\documentclass[letterpaper, 10 pt, conference]{ieeeconf}  

\IEEEoverridecommandlockouts                             
\usepackage{amsmath, amssymb, mathrsfs}
\usepackage[utf8]{inputenc}
\usepackage[T1]{fontenc}
\usepackage{graphicx}
\usepackage[table]{xcolor}
\usepackage{colortbl}
\usepackage{caption}
\usepackage{tabularx}
\usepackage{amssymb}
\usepackage{array, xcolor, colortbl}
\usepackage{multirow}
\usepackage{tikz}
\usepackage{algorithm}
\usepackage{algpseudocode}
\definecolor{green_arrow}{RGB}{0, 128, 0}
\definecolor{red_arrow}{RGB}{255, 0, 0}
\definecolor{lightgray}{gray}{0.9}
\definecolor{lightblue}{rgb}{0.75, 0.85, 0.95}
\definecolor{lightgreen}{rgb}{0.75, 0.95, 0.75}
\definecolor{lightyellow}{rgb}{0.95, 0.95, 0.75}

\title{\LARGE \bf
PAM-UNet: Shifting Attention on Region of Interest in Medical Images}

\author{Abhijit Das$^{1}$, Debesh Jha$^{1}$, Vandan Gorade$^{1}$,  Koushik Biswas$^{1}$, Hongyi Pan$^{1}$, Zheyuan Zhang$^{1}$, \\ Daniela P. Ladner$^{2}$,  Yury Velichko$^{1}$, Amir Borhani$^{1}$, and Ulas Bagci$^{1}$
\thanks{*This work is supported by the NIH funding: R01-CA246704, R01-CA240639, U01-DK127384-02S1, and U01-CA268808.}
\thanks{$^{1}$A. Das, D. Jha, V. Gorade, K. Biswas, H. Pan,  Z. Zhang and U. Bagci are with the Machine \& Hybrid Intelligence Lab, Department of Radiology, Northwestern University, USA. }%
\thanks{$^{2}$D.P. Ladner is with Comprehensive Transplant Center, Feinberg School of Medicine, Northwestern University, USA  \& Division of Transplantation, Department of Surgery, Northwestern Medicine, USA.}   
}

\begin{document}
\maketitle
\thispagestyle{empty}
\pagestyle{empty}
\begin{abstract}
Computer-aided segmentation methods can assist medical personnel in improving diagnostic outcomes. While recent advancements like UNet and its variants have shown promise, they face a critical challenge: balancing accuracy with computational efficiency.  Shallow encoder architectures in UNets often struggle to capture crucial spatial features, leading in inaccurate and sparse segmentation. To address this limitation, we propose a novel \underline{P}rogressive \underline{A}ttention based \underline{M}obile \underline{UNet} (\underline{PAM-UNet}) architecture. The inverted residual (IR) blocks in PAM-UNet help maintain a lightweight framework, while layerwise \textit{Progressive Luong Attention} ($\mathcal{PLA}$) promotes precise segmentation by directing attention toward regions of interest during synthesis. Our approach prioritizes both accuracy and speed, achieving a commendable balance with a mean IoU of 74.65 and a dice score of 82.87, while requiring only 1.32 floating-point operations per second (FLOPS) on the Liver Tumor Segmentation Benchmark (LiTS) 2017 dataset. These results highlight the importance of developing efficient segmentation models to accelerate the adoption of AI in clinical practice.
\end{abstract}
\section{\textbf{Introduction}}
The growing demand for precise segmentation in medical images has led to the development of complex models like HRNet~\cite{zhang2021refined}, DeepLabv3+~\cite{yu2022lightweight}, AttUNet~\cite{wang2021attu}, and FANet~\cite{tomar2022fanet} which rarely fit on edge devices. While adopting shallower backbones in UNet architectures, such as Efficient-UNet~\cite{baheti2020eff} and Mobile-UNet~\cite{jing2022mobile}, reduces computational requirements, it frequently comes at the cost of segmentation quality. These models often struggle to differentiate indistinguishable peripheral and occluded patterns, and handle the large inter-patient variation of anomalies, due to the inability of lightweight encoding modules to generate significantly rich feature maps.

Fully convolutional CNNs (FCNNs)~\cite{cheng2020fully} and UNet~\cite{ronneberger2015u} have paved the way in biomedical image segmentation. UNet inspired networks, like DoubleUNet~\cite{jha2020doubleu}, ResNet-UNet~\cite{diakogiannis2020resunet}, Attention-UNet~\cite{wang2021attu} excel in organ segmentation and lesion detection. However, extensive down-sampling using deeper backbones results in spatial information loss. Mobile-UNet aims to mitigate overfitting and optimize computation by employing a pre-trained MobileNetV2~\cite{dong2020mobilenetv2} as the encoder, but the limited capacity for capturing fine-grained features results in spatial context loss. 

This motivated us to propose \textbf{PAM-UNet}, a new segmentation architecture based on depth-wise separable convolution and a novel \textit{Progressive Luong Attention} ($\mathcal{PLA}$) mechanism. $\mathcal{PLA}$ selectively incorporates long-range dependencies from shallow features to the final segmentation mask, guided by residual cues from the encoder. This targeted approach, unlike conventional attention mechanisms, focuses only on relevant regions. This synergistic design, combined with Mobile convolutions (MB Conv) and layerwise Luong Loss, significantly reduces FLOPs while enhances feature representation. Our key contributions are summarized as follows:

\begin{itemize}
    \item \textbf{Novel  architecture}: We introduce a novel architecture, \textbf{\textit{PAM-UNet}}, that combines mobile convolution blocks along with $\mathcal{PLA}$ mechanism for precise medical image segmentation with heightened efficiency.
    \item \textbf{Advanced attention mechanism}: Our study show superior performance of $\mathcal{PLA}$ attention over self-and cross-attention mechanisms.
    \item \textbf{Detailed interpretation}: We provide a detailed interpretation to show how PAM-UNet is different from and superior to existing baselines using center kernel alignment (CKA)~\cite{wang2020centered}. 
    \item \textbf{Extensive evaluation}: We have throughly evaluated PAM-UNet on two common benchmarks: LiTS-2017~\cite{bilic2023liver} and Kvasir-SEG~\cite{jha2020kvasir} dataset. The results show that PAM-UNet achieves comparable precision with $30x$ less computational overhead.
\end{itemize}

\begin{figure}[t]
    \centering
    \includegraphics[width=1\linewidth]{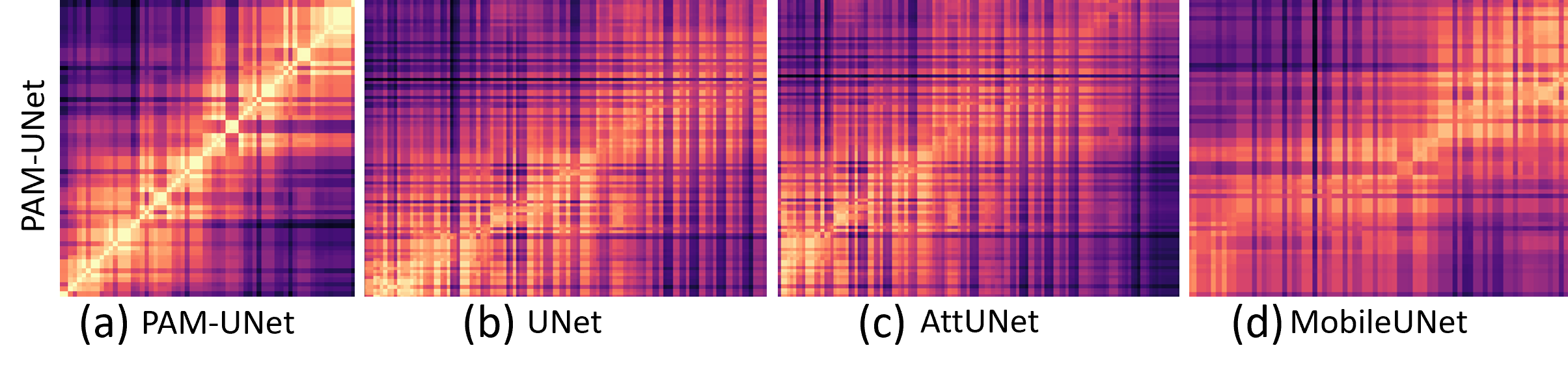}
    \captionsetup{skip=1pt}
    \caption{\textbf{CKA plot illustrating that PAM-UNet learns from the very beginning.} (a) PAM-UNet have highly similar structure throughout the model, (b) and (d) A large number of lower layers in the UNet and MobileUNet are similar to smaller set of the lower PAM-UNet layers and (c) PAM-Net takes a few layers lesser than AttUNet to learn the similar representations during training.}
    \label{fig:cka-comparisons}
    \vspace{-5mm}
\end{figure}


\vspace{-2mm}
\section{\textbf{Preliminaries}}
\noindent \textbf{Vanilla vs Depth-wise separable convolution~\cite{chollet2017xception}:} Traditional convolution performs a linear combination of input values and kernels across all dimensions and channels. In contrast, depth-wise separable convolution decomposes the convolution operations into depth-wise and point-wise convolutions separately and performs:
\begin{figure*}[!ht]
    \centering
    \includegraphics[width=0.97\linewidth]{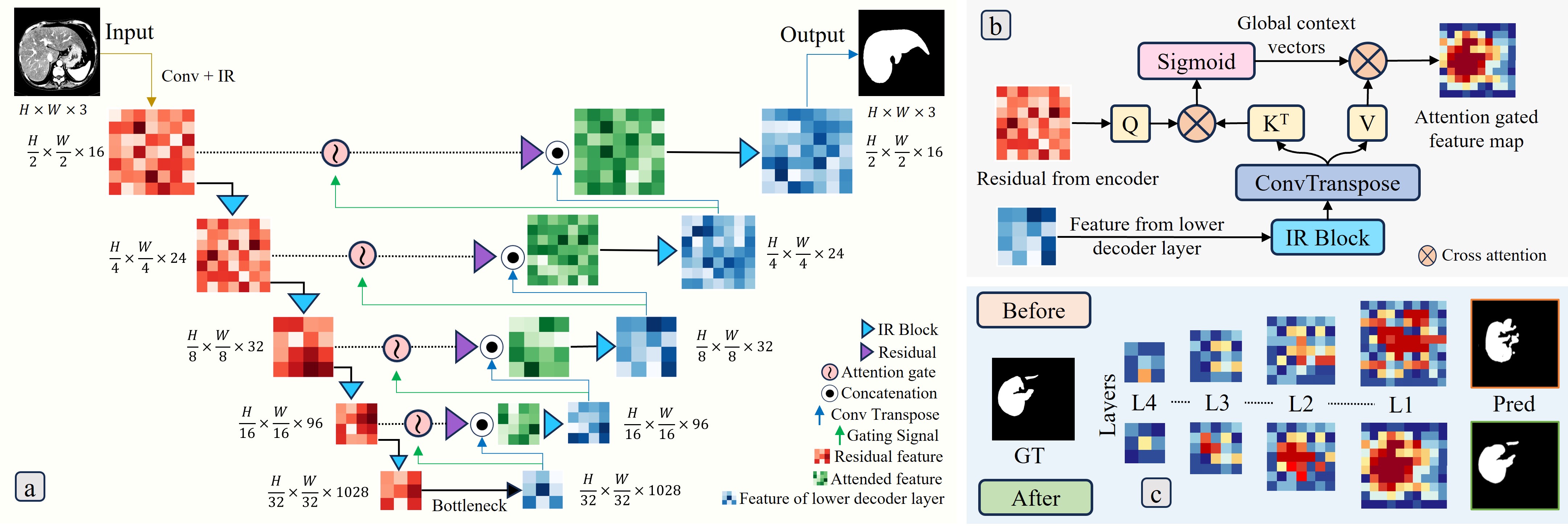}
    \caption{Here we present (a) Holistic architecture of proposed PAM-UNet, (b) Diagram of working of one unit of attention gate ($\mathcal{PLA}$) and (c) Features extracted from intermediate layers of PAM-UNet showing the effectiveness of proposed $\mathcal{PLA}$. Predicted mask before and after applying $\mathcal{PLA}$ is significantly different and proposed method produces precise segmentation.}
    \label{fig:model}
\end{figure*}
\begin{equation}
\label{eq:depthwise_pointwise_compact}
\begin{aligned}
&DSConv = \mathcal{DP}(\mathbf{I}, \mathbf{K}_{\mathrm{d}}, \mathbf{K}_{\mathrm{p}}) = \\ &\underbrace{\sum_{m,n} \mathbf{I}(i+m, j+n, k) \ast \mathbf{K}_{\mathrm{d}}(m, n, k)}_{\text{Depth-wise ($\mathcal{D}$)}} \, \underbrace{\sum_{k} \ast \mathbf{K}_{\mathrm{p}}(k, l)}_{\text{Point-wise  ($\mathcal{P}$)}}
\end{aligned}
\end{equation}
where $\mathcal{DP}(\cdot)$ performs depth-wise and point-wise convolution, $\mathbf{I}$ is the input tensor, $\mathbf{K}_{\mathrm{d}}$ and $\mathbf{K}_{\mathrm{p}}$ are depth-wise and point-wise kernels, respectively. $\sum_{m,n}$ works over all spatial dimensions \( m \) and \( n \).

Depth-wise separable convolution $DSConv$ reduces computational cost by applying separate kernels to individual input channels, making it more efficient than vanilla convolution.

\noindent \textbf{Self vs Cross~\cite{vaswani2017attention} vs Luong attention~\cite{luong2015effective}:} Self-attention captures intra-sequence relationships via the weighted sum of input features by learnable parameters and activation (\(\sigma\)). Cross-attention extends this to inter-sequence relationships. In contrast, Luong attention computes the context vector \(c_t\) for each decoder time step \(t\) as a weighted sum of encoder hidden states \(h_i\):
\begin{equation}
c_t = \sum_{i=1}^{T} \alpha_{t,i} \cdot h_i
\end{equation}
The attention weights \(\alpha_{t,i}\) are determined by a $\sigma$ function applied to a score function \(score(s_t, h_i)\), measuring the similarity between the decoder hidden state \(s_t\) and encoder hidden states:
\begin{equation}
\alpha_{t,i} = \frac{\exp(score(s_t, h_i))}{\sum_{j=1}^{T} \exp(score(s_t, h_j))}
\end{equation}

\section{\textbf{Proposed Method}}  
The proposed PAM-UNet adopts a U-shaped architecture with dedicated down-sampling and up-sampling arms. Each arm employs mobile convolution blocks equipped with layer-wise additive luong attention to capture relevant features at different scales. A detailed description is mentioned below:

\subsection {\textbf{Decoding the Lightweight U-shaped backbone}}
The proposed PAM-UNet model, illustrated in Fig.~\ref{fig:model}a, comprises an encoding arm and a decoding arm with skip connections between corresponding layers. The encoding arm incorporates MB Conv, IR bottleneck, and a convolutional layer for dimensionality reduction. The decoding arm mirrors the encoding arm's structure, utilizing layerwise Luong attention on skip connections. Unlike UNet's bottleneck block, MBConv employs an IR bottleneck and performs \textit{ConvTranspose} instead of \textit{up-sampling} during synthesis. The attention mechanism focuses on relevant information in the encoder's residual features. The final layer by through a ConvTranspose generates the segmentation mask.
\subsection {\textbf{Progressive Luong Attention}}
$\mathcal{PLA}$ is designed to capture hierarchical context dependencies. While the standard Luong attention attends to the information from a single layer, $\mathcal{PLA}$ refines this process by considering a progressive aggregation of information from multiple layers. As shown in Fig.\ref{fig:model}b, feature map from lower decoding layer is passed through a IR Block and ConvTranspose splits into \textit{Key} and \textit{Value} and the residual coming from the current encoding block is used as \textit{Query}. Given query (\(Q\)) and key (\(K\)) vectors, the attention scores (\(A\)) are calculated as:
\begin{equation}
    A = \sigma\left(\frac{Q \cdot K^T}{\sqrt{\text{key\_dimension}}}\right)
\end{equation}
The attention weights (\(W\)) are obtained by applying $\sigma$ to the attention scores, and attended values (\(V\)) are computed by:
\begin{equation}
    W = {\sigma}(A) \quad \text{and} \quad V_{\text{attended}} = W \cdot V
\end{equation}
\begin{table*}[!ht]
    \centering
    \caption{Results comparison of PAM-UNet with other popular benchmarking algorithms. Here, UNet variants with an enhanced attention mechanism are shown in \textcolor{orange!85}{orange} and variants with improved backbone architecture is shown in \textcolor{cyan!85}{cyan}.}
  
    \renewcommand{\arraystretch}{1.5} 
    \begin{tabularx}{\linewidth}{l|*{3}{>{\centering\arraybackslash}X}|*{3}{>{\centering\arraybackslash}X}|>{\centering\arraybackslash}X}
    \hline
        \rowcolor{blue!10}
        & \multicolumn{3}{c}{\textbf{LiTS}} & \multicolumn{3}{c}{\textbf{Kvasir-SEG}} & \\ \cline{2-7}
        \rowcolor{blue!10} \textbf{Model} & \textbf{Dice (\%)} \(\uparrow\) & \textbf{mIoU (\%)} \(\uparrow\) & \textbf{Recall (\%)} \(\uparrow\) & \textbf{Dice (\%)} \(\uparrow\) & \textbf{mIoU (\%)} \(\uparrow\) & \textbf{Recall (\%)} \(\uparrow\) &\textbf{FLOPS} \(\downarrow\)\\ \hline
        U-Net (ResNet50)~\cite{ronneberger2015u} & 78.02 & 69.00 & 87.20 & \textcolor{red}{\textbf{83.57}} & 76.8 & 84.26 & 35.00 \\
        \rowcolor{orange!11}
        DeepLabv3+~\cite{Chen_2018_ECCV} & 82.34 & 73.59 & 91.11 & 81.72 & 78.36 & 85.24 & 27.90 \\
        \rowcolor{orange!11}
        \rowcolor{orange!11}
        Mobile-UNet~\cite{jing2022mobile} & 71.04 & 62.93 & 80.23 & 76.45 & 71.65 & 78.43 & \textbf{0.85} \\
        \rowcolor{orange!11}
        FCN8~\cite{cheng2020fully} & 76.86 & 67.65 & 82.5 & 68.98 & 64.52 & 73.50 & 24.90 \\
        \rowcolor{orange!11}
        HRNet~\cite{zhang2021refined} & 76.21 & 72.85 & \textbf{92.34} & 80.46 & 77.87 & \textbf{87.43} & 11.70 \\
        \rowcolor{green!15}
        \rowcolor{cyan!15}
        ResUNet~\cite{diakogiannis2020resunet} & 74.09 & 64.45 & 78.98 & 65.21 & 59.41 & 70.45 & 14.82 \\
        \rowcolor{cyan!15}
        AttUNet~\cite{wang2021attu} & \textcolor{red}{\textbf{82.38}} & \textcolor{red}{\textbf{74.32}} & 76.95 & 83.24 & \textbf{79.91} & \textcolor{red}{\textbf{87.16}} & 35.35 \\
        \rowcolor{green!15}
        \textbf{PAM-UNet (Ours)} & \textbf{82.87} & \textbf{74.65} & \textcolor{red}{\textbf{92.14}} & \textbf{84.8} & \textcolor{red}{\textbf{78.40}} & 86.63 & \textcolor{red}{\textbf{1.32}}\\\hline
    \end{tabularx}
    \label{tab:model-performance}
\end{table*}


\subsection {\textbf{Loss function}}
During back-propagation, gradients from irrelevant regions are effectively \textit{down-weighted}, prioritizing updates of relevant spatial regions. Assuming $\hat{Y}$ represents the predicted segmentation map and $Y$ represents the ground truth segmentation map, the segmentation loss ($\mathcal{L}_{\text{seg}}$) is:
\begin{equation}
    \mathcal{L}_{\text{seg}} = -\frac{1}{N} \sum_{i=1}^{N} \left( Y_i \cdot \log(\hat{Y}_i) + (1 - Y_i) \cdot \log(1 - \hat{Y}_i) \right)
\end{equation}

\noindent The additional objective of attention mechanism is further enhanced by Attention Regularization to mitigate over-attentiveness that puts attention on both background and anomaly together. This is defined as:
\begin{equation}
\mathcal{R}(\mathbf{A}) = \frac{1}{N} \sum_{i=1}^{N} (\mathbf{A}_i - \bar{A})^2
\end{equation}
\begin{equation}
    \mathcal{L}_{\text{total}} = \mathcal{L}_{\text{seg}} +  \lambda \cdot \mathcal{L}_{\text{$\mathcal{R}$}}
\end{equation}
This formulation aims to train the model by encouraging the attention mechanism to distribute its focus more evenly.

\section{\textbf{Experiments}}
\noindent \textit {\textbf{Evaluation datasets:}}
We used two publicly available datasets, namely \textit{Liver Tumor Segmentation (LiTS)-2017}~\cite{bilic2023liver} and \textit{Kvasir-SEG}~\cite{jha2020kvasir}, each representing distinct imaging modalities (CT and Colonoscopy respectively) for comprehensive evaluation of performance and compatibility of our model. The LiTS dataset contains 201 CT images of the abdomen, but only the training dataset (130 scans) is made publicly available, so we have split the \textit{train} set of LiTS into \textit{train} and \textit{test}. Kvasir-SEG comprises 1000 colonoscopy images. We used a split of 80\% for training and the remaining for testing and validation. 

\noindent \textit{\textbf{Metrics:}}
Aligning with segmentation challenge evaluation standards, we adopt the \textit{Dice Similarity Coefficient (Dice), mean Intersection Over Union (mIoU) and Recall} to measure the quality of predicted segmentation masks and to evaluate computation efficiency we have picked \textit{FLOPS}. To further compare the representations learned by PAM-UNet and other models across different layers we use CKA plots. CKA goes beyond comparing individual activations and instead compares how data points relate to each other within the representation space.

\noindent \textit{\textbf{Implementation Details:}}
Our models were implemented using PyTorch and trained on a NVIDIA RTX-A4000 GPU with 16GB of memory. The input image size was set to $128 \times 128$, and training utilized a batch size of 8 with a learning rate of 0.01. We employed the SGD optimizer with momentum 0.9 and weight decay of 0.0001 for optimization. Data augmentations such as flipping and rotating were applied during training to enhance model generalization.

\noindent \textit{\textbf{Architecture Configuration:}}
PAM-UNet employs MBConv blocks and $\mathcal{PLA}$ mechanism within a mirrored U-shaped architecture. It incorporates DSConv for flexible expansion factors and strides (1 to 2), along with optional deconvolution in IR blocks. To mitigate overfitting, attention regularization is applied. Key hyperparameters include an expansion factor of 6, stride variation from 1 to 2, and an attention regularization weight of 0.01.

\noindent \textit{\textbf{Techniques for Comparison:}} We compare PAM-UNet against UNet and six other variants: MobileUNet~\cite{jing2022mobile}, DeepLabV3+~\cite{Chen_2018_ECCV}, FCN-8~\cite{cheng2020fully}, HRNet~\cite{zhang2021refined}, ResUNet~\cite{diakogiannis2020resunet}, and AttUNet~\cite{wang2021attu}. The first four variants utilize diverse backbone implementations, while the remaining two incorporate attention gating methodologies. This separation aims to highlight the efficacy of a mobile convoluted backbone and the proposed $\mathcal{PLA}$ gating in PAM-UNet with all fairness.
\begin{figure*}
    \centering
    \includegraphics[width=\linewidth]{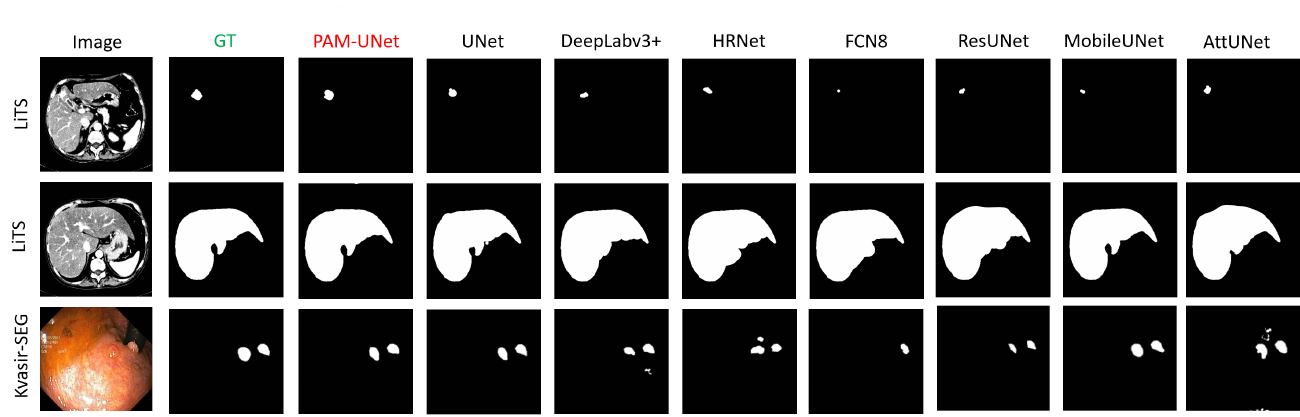}
    \caption{Segmentation masks generated by \textcolor{red}{PAM-UNet (Ours)} and all the baselines highlight the capability justifying quantitative results. PAM-UNet segments both \textit{liver} and \textit{tumour} precisely detecting boundaries better than MobileUNet and AttUNet. DeepLabv3+, FCN8 and ResUNet due to overfitting takes peripheral regions in segmentation mask also while segmenting liver.}
    \label{fig:model-segmentation-performnces}
\end{figure*}
\section{\textbf{Results}}
\subsection{\textbf{Quantitative results}}
From Table \ref{tab:model-performance}, we note that proposed PAM-UNet outperforms both backbone and attention mechanism based baselines. Overall, PAM-UNet secures the top spot in Dice score across both datasets, surpassing complex models like U-Net (ResNet50), DeepLabv3+, and FCN8. It even challenges the impressive performance of attention-based variants like AttUNet, achieving comparable mIoU and Dice scores while maintaining significantly lower computational cost of 1.32 FLOPS (vs. 35.35 FLOPS). This efficiency advantage becomes even more evident when compared to MobileUNet, where PAM-UNet achieves substantial improvements in Dice score of 11.83\% on LiTS and 8.35\% on Kvasir-SEG and mIoU of 11.72\% on LiTS and 6.75\% on Kvasir-SEG, while requiring only marginally more FLOPs (1.32 vs. 0.85). 

\begin{table}[!h]
\centering
\caption{Effect of \textcolor{cyan!85}{mobile blocks} and different \textcolor{purple!85}{attentions}.}
\renewcommand{\arraystretch}{1.5} 
\begin{tabularx}{\linewidth}{lXXX} 
\hline
\rowcolor{blue!13}\textbf{Method} & \textbf{Dice} \(\uparrow\) & \textbf{mIoU} \(\uparrow\) & \textbf{FLOPS} \(\downarrow\)\\
\hline
\rowcolor{cyan!8}w/ Mob (M) Encoder only & 71.04 & 62.93 & 0.85\\
\rowcolor{cyan!12}w/ M Encoder-decoder (MED) & 75.89 & 71.28 & \textbf{0.72}\\
\rowcolor{purple!8}MED + Self attention & 78.32 & 73.56 & 0.98\\
\rowcolor{purple!8}MED + Cross attention & 81.26 & 74.44 & 1.24\\
\rowcolor{purple!8}MED + Additive attention & 79.40 & 74.34 & 1.17\\
\rowcolor{green!17} \textbf{MED+ $\mathcal{PLA}$ (Ours)} & \textbf{82.87} & \textbf{74.65} & 1.32\\
\hline
\end{tabularx}
\label{tab:ablation}
\vspace{-5mm}
\end{table}

\subsection{\textbf{Qualitative results}}
As shown in Fig.~\ref{fig:model-segmentation-performnces} PAM-UNet consistently identifies regions of interest with more refined boundaries, outperforming baselines. PAM-UNet demonstrates precise predictions of lesions, even in cases involving varying small locations, sizes, and modality. Furthermore, it efficiently suppresses irrelevant information, including background elements.

\subsection{\textbf{Ablation study}}
Table~\ref{tab:ablation} summarizes the ablation study, revealing the $\mathcal{PLA}$ as the most effective mechanism, resulting in significant enhancements in mIoU and Dice scores. Solely relying on self-attention may not fully exploit contextual information from other parts of the input, resulting in suboptimal segmentation performance. Although cross attention improves segmentation performance compared to self-attention alone, the increased computational cost may limit its practicality for resource-constrained scenarios. Whereas, additive attention introduces a learnable transformation matrix to compute attention scores, offering flexibility in capturing complex relationships between input tokens. However, the performance improvement is modest compared to cross attention.

Our proposed method, \(\mathcal{PLA}\) effectively captures long-range dependencies while incorporating positional information, allowing the model to better understand spatial relationships within the input. Although it incurs a slightly higher computational cost compared to other attention mechanisms, the substantial performance gains justify its adoption. By striking a balance between accuracy and computational efficiency, \(\mathcal{PLA}\) emerges as the preferred choice for semantic segmentation tasks.

\subsection{\textbf{Discussion}}
Experimental results and ablation studies shows that the significant improvement in Dice and mIoU scores stems primarily from the incorporation of $\mathcal{PLA}$. Fig.~\ref{fig:model}(c) demonstrates this effect by illustrating how PLA alters feature map representations at each decoding layer. Furthermore, Fig.~\ref{fig:cka-comparisons}(a) highlights how the symmetric structure of PAM-UNet ensures equal learning opportunities across all layers. As shown in Fig.~\ref{fig:cka-comparisons}(b, c and d), PAM-UNet exhibits greater learning efficiency from initial layers compared to UNet, AttUNet, and MobileUNet. This implies that PAM-UNet achieves comparable representation learning with a very shallow light-weight encoder. 
\begin{figure}[!t]
    \centering
    \includegraphics[width=0.8\linewidth]{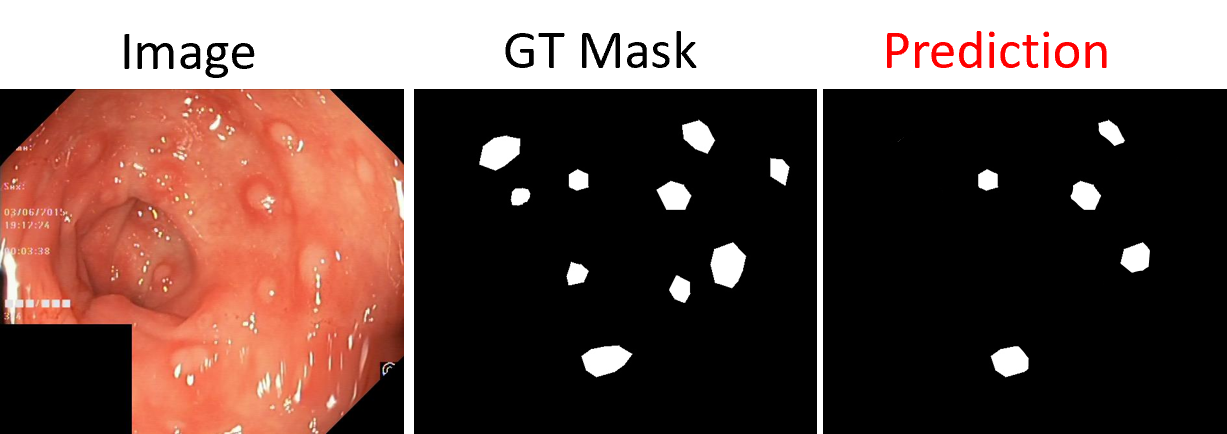}
    \caption{\textbf{Corner case showing where PAM-UNet failed} to detect all 10 polyps, which is a rare case in the Kvasir-SEG dataset (No training image contains more than 5 polyps).}
    \label{fig:corner-case-pamunet}
\end{figure}

\subsection{\textbf{Limitations}}
The PAM-UNet model, while showing promising performance, has limitations. As shown in Fig.~\ref{fig:corner-case-pamunet}, PAM-UNet failed to segment precisely. The investigation of the failure cases suggests that no image in training data consists of more than five polyps. Leveraging the training data with a variety of training samples can mitigate the issue. We also recognize the recent rise of Transformer based segmentation architectures. However, it should be noted that these architectures become very successful when combined with UNet-style architectures, indicating the necessity of U-shaped architectures and/or encoder-decoder styles in successfully locating the boundary of objects. Since our focus in this study is to provide an effective segmentation engine with lightweight architecture, and since Transformer based architectures are currently costly in computational burden, we focus on the architecture stemming from lightweight and U-shaped architectures. In future work, we will extend our strategy by encoding linear Transformers in the same problem setting. Also we will evaluate proposed method on 3D segmentation task over variety of modalities.
\section{\textbf{Conclusion}}
In summary, we present a novel medical image segmentation architecture, PAM-UNet, that utilizes mobile convolutions and $\mathcal{PLA}$ to achieve state-of-the-art biomedical image segmentation performance against the baselines while maintaining a low computational cost of only 1.32 FLOPS. Our experiments demonstrate superior performance on LiTS and Kvasir-SEG benchmarks, with ablation studies confirming the crucial role of progressive attention in capturing long-range dependencies. Qualitative analysis further showcases PAM-UNet's ability to learn relevant representations from initial layers. While acknowledging limitations in specific scenarios, we see PAM-UNet as an efficient solution for accurate and efficient medical image segmentation, paving the way for future advancements in the field.


\bibliographystyle{plain}
\bibliography{main}
\end{document}